\documentclass[aps,pra,twocolumn,amsmath,amssymb]{revtex4}
\usepackage{graphicx}
\begin{document}

\title{Controlled Unitary Operation between Two Distant Atoms}
\author{Jaeyoon Cho}
\affiliation{Division of Optical Metrology, Korea Research Institute of Standards and Science, Daejeon 305-340, Korea}
\author{Hai-Woong Lee}
\affiliation{Department of Physics, Korea Advanced Institute of Science and Technology, Daejeon 305-701, Korea}
\date{\today}
\begin{abstract}
We propose a scheme for implementing a controlled unitary gate between two distant atoms directly communicating through a quantum transmission line. To achieve our goal, only a series of several coherent pulses are applied to the atoms. Our scheme thus requires no ancilla atomic qubit. The simplicity of our scheme may significantly improve the scalability of quantum computers based on trapped neutral atoms or ions.
\end{abstract}
\maketitle

\newcommand{\bra}[1]{\left<#1\right|}
\newcommand{\ket}[1]{\left|#1\right>}
\newcommand{\ketbra}[2]{\left|#1\right>\left<#2\right|}
\newcommand{\abs}[1]{\left|#1\right|}


\section{Introduction}

A quantum optical system utilizing trapped neutral atoms or ions for qubits is one of promising candidates for implementing a quantum computer \cite{m02}. Actually, there have been numerous theoretical \cite{cz95,cz00} and experimental \cite{shr03} achievements showing the positive prospects for it. The number of qubits in such a system is, however, obviously limited by the size of the trapping structure, while one of the essential factors for a useful quantum computer is the scalability. This difficulty could be overcome by connecting partially implemented quantum computation nodes to form a quantum network.

For any unitary operation for the whole quantum network to be possible, controlled unitary operations between two nodes should be performed as well as local unitary operations at each node \cite{nc00}. There have been various ways for doing it by means of one or more ancilla qubits \cite{m02}. The underlying idea is to use ancilla qubits to transfer the quantum information between two nodes and perform local two-qubit operations at the nodes so that the overall process in effect results in the desired global two-qubit operation. One method to accomplish the task is to shuttle ancilla qubits or a quantum node itself physically to a particular position where local interaction between an ancilla qubit and a quantum node is possible \cite{cz00}. This method, however, cannot be directly applicable to neutral atom quantum computers. Another method feasible for neutral atom quantum computers as well is to exploit a photon-mediated interaction between two nodes, such as in entanglement generation \cite{fzd03}, quantum state transfer \cite{czk97}, and quantum teleportation \cite{bkp99}. On the other hand, there have also been a scheme in which no ancilla qubit is involved \cite{sm98}. The scheme, however, uses a quantum interferometer and the complex atomic structure of several hyperfine levels instead.

In this work, we introduce a simple scheme to do a controlled unitary operation between two distant atoms. A common quantum communication setup \cite{czk97}, in which two atoms each trapped in an optical cavity directly communicate through a quantum transmission line such as an optical fiber connecting the two cavities, is considered. In contrast to earlier methods, no ancilla atomic qubit is involved in our scheme and the gate operation is done by a simple coherent process.


\section{The Scheme}


\begin{figure}[b]
\includegraphics[width=0.40\textwidth]{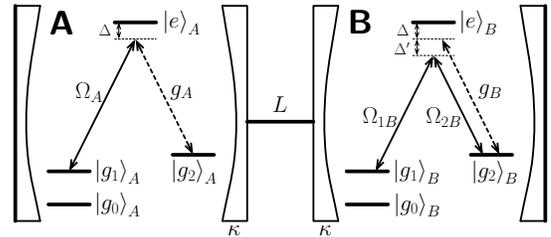}
\caption{The schematic representation of the controlled phase shift gate between two distant atoms. Two atoms are each trapped in an optical cavity and the two cavities are connected with a fiber. Each solid arrow represents a transition by the classical field and each dotted arrow a transition by the cavity mode. A qubit is encoded in two atomic ground levels $\ket{g_0}$ and $\ket{g_1}$.}
\label{fig:scheme}
\end{figure}

The schematic representation of our scheme is depicted in Fig.~\ref{fig:scheme}. Atom A and B are trapped in cavity A and B, respectively, and two cavities are connected through an optical fiber of length $L$. The decay rate of the cavity is $\kappa$ and the spontaneous emission rate of the atom is $\gamma$. The outside mirror, i.e., the mirror on the side not connected to the fiber, of each cavity is assumed to be of 100\% reflectivity. Each solid arrow represents a transition by a classical field of Rabi frequency $\Omega_i(t)$ ($i=A,1B,2B$), and each dotted arrow represents a transition by a cavity mode of atom-cavity coupling rate $g_i$ ($i=A,B$). We assume the Lamb-Dicke limit \cite{dkk03}, thus $g_i$ is assumed to be a constant. Each transition is detuned by an amount of $\Delta$ or $\Delta+\Delta'$ as shown in Fig.~\ref{fig:scheme}. A qubit is represented by two ground hyperfine levels $\ket{g_0}$ and $\ket{g_1}$. $\ket{g_0}$ does not participate in the transition.

The desired two qubit operation is the controlled phase shift operation, which is accomplished in three steps as the following:
\begin{widetext}
\begin{equation}
\left(\begin{matrix}
\ket{g_{0}}_{A}\ket{g_{0}}_{B} \\ \ket{g_0}_A\ket{g_1}_B \\ 
\ket{g_1}_A\ket{g_0}_B \\ \ket{g_1}_A\ket{g_1}_B
\end{matrix}\right)
\underrightarrow{\text{ 1st step }}
\left(\begin{matrix}
\ket{g_0}_A\ket{g_0}_B \\ \ket{g_0}_A\ket{g_2}_B \\ 
\ket{g_1}_A\ket{g_0}_B \\ \ket{g_1}_A\ket{g_2}_B
\end{matrix}\right)
\underrightarrow{\text{ 2nd step }}
\left(\begin{matrix}
\ket{g_0}_A\ket{g_0}_B \\ \ket{g_0}_A\ket{g_2}_B \\ 
\ket{g_1}_A\ket{g_0}_B \\ e^{i\phi}\ket{g_1}_A\ket{g_2}_B
\end{matrix}\right) 
\underrightarrow{\text{ 3rd step }}
\left(\begin{matrix}
\ket{g_0}_A\ket{g_0}_B \\ \ket{g_0}_A\ket{g_1}_B \\ 
\ket{g_1}_A\ket{g_0}_B \\ e^{i\phi}\ket{g_1}_A\ket{g_1}_B
\end{matrix}\right).
\label{eq:step}
\end{equation}
\end{widetext}
In the first step, only the state $\ket{g_1}_B$ is transferred to $\ket{g_2}_B$ while other states remain unchanged. It is done with high precision by the well-known technique of adiabatic passage \cite{bts98}. For this step, two classical fields of Rabi frequency $\Omega_{2B}(t)$ and $\Omega_{1B}(t)$ are applied adiabatically to atom B in order. Here, detuning parameter $\Delta'$ has to be much larger than atom-cavity coupling rate $g_{B}$ ($\Delta'\gg g_B$) for no cavity photon to be generated during the adiabatic passage process. The third step is simply the inverse of the first step, and is also achieved by adiabatic passage. The most important and nontrivial part of our scheme is the second step, in which only the state $\ket{g_1}_A\ket{g_2}_B$ acquires a phase $\phi$ while other states remain unchanged. In the remainder of this paper, we concentrate on explaining the second step of operation~(\ref{eq:step}).

In the second step, a classical field of Rabi frequency $\Omega_A(t)$ is applied to atom A adiabatically, whereas both the classical fields of Rabi frequencies $\Omega_{1B}$ and $\Omega_{2B}$ for atom B are turned off. Let us assume that $\Omega_{A}(t)$ has a Gaussian form:
\begin{equation}
\Omega_A(t)=\Omega_{0}\exp\left[-\left(\frac{t-t_c}{\Delta_t}\right)^2\right].
\label{eq:rabi}
\end{equation}
If the initial state of atom A is $\ket{g_{0}}_{A}$, this operation has no effect on the system. If atom A is initially in state $\ket{g_1}_A$, however, this operation transfers the population to $\ket{g_2}_A$, during which a single photon is generated in cavity A and emitted out of the cavity \cite{dkk03,khr02}. The time width $\Delta_f$ of the emitted photon pulse is of order $\Delta_t$. With that in mind, we investigate the system in two regimes.


\section{Short-Distance Regime}

First, we consider the short-distance regime in which the interaction time between two distant cavities is sufficiently short so that the whole system can be regarded to remain in a steady state at all times. This regime is represented by the following condition:
\begin{equation}
\Delta_t\gg 2L/c,
\end{equation}
where $c$ is the speed of light. In this case, the whole system can be treated within the context of adiabatic theorem.

If the system is initially in state $\ket{g_{1}}_{A}\ket{g_{0}}_{B}$, the Hamiltonian in the rotating frame is written as
\begin{equation}
\begin{split}
H&=H_{A}+H_{B}+H_{C}, \\
H_{A}&=(\Delta-i\gamma/2)\ket{e}_{A}\bra{e} \\
 &\quad+[\Omega_{A}(t)\ket{e}_{A}\bra{g_{1}}+g_{A}a_{A}\ket{e}_{A}\bra{g_{2}}+h.c.],\\
H_{B}&=0, \\
H_{C}&=\sum_{n=-\infty}^{\infty} n\delta\omega\, c_{n}^{\dagger}c_{n} \\
&\quad+i\kappa'\sum_{n=-\infty}^{\infty}[a_{A}^{\dagger}c_{n}+(-1)^{n}a_{B}^{\dagger}c_{n}-h.c.],
\end{split}
\label{eq:hamiltonian1}
\end{equation}
where $H_{A}$, $H_{B}$, and $H_{C}$ are the Hamiltonians for atom A, atom B, and the fiber, respectively, $a_{A}$ ($a_{B}$) is the field operator for cavity A (cavity B),  $c_{n}$ is the field operator for the $n$th fiber mode, $\delta\omega=\pi c/L$ is the frequency difference between two adjacent fiber modes, and $\kappa'\equiv\sqrt{\frac{\kappa\delta\omega}{2\pi}}$ is the effective cavity decay rate. The factors $(-1)^{n}$ are introduced to model the phase difference $\pi$ between two ends of the fiber for every second modes. Hamiltonian $H_{A}$ has the dark state
\begin{equation}
\ket{D{(t)}}_{A}=\cos\theta_{A}(t)\ket{g_{1},0}_{A}-\sin\theta_{A}\ket{g_{2},1}_{A},
\end{equation}
where $\theta_{A}(t)$ is given by $\tan\theta_{A}(t)=\Omega_{A}(t)/g_{A}$. Here, we represent a state of the atom-cavity system as $\ket{x,n}$ where $x$ is the atomic state and $n$ is the cavity photon number. The dark state of the total Hamiltonian $H$ is also derived as
\begin{equation}
\begin{split}
&\ket{D(t)}_{AB}\propto\ket{D(t)}_{A}\ket{g_{0},0}_{B}\ket{0}_{C} \\
&\quad+\sin\theta_{A}(t)\ket{g_{2},0}_{A}\ket{g_{0},1}_{B}\ket{0}_{C} \\
&\quad-i\kappa'\!\sum_{n=-\infty}^{\infty}\frac{2\sin\theta_{A}(t)}{(2n+1)\delta\omega}\ket{g_{2},0}_{A}\ket{g_{0},0}_{B}\ket{1_{2n+1}}_{C},
\end{split}
\label{eq:dark}
\end{equation}
where $\ket{0}_{C}$ and $\ket{1_{j}}_{C}$ denote the vacuum fiber and one photon in the $j$th fiber mode, respectively. From this expression, it is clear that after the classical pulse operation given by Eq.~(\ref{eq:rabi}) the system just returns to its initial state since both the initial and the final values of $\Omega_{A}$ are 0, i.e., $\theta_{A}(initial)=\theta_{A}(final)=0$. During this operation, the dark state acquires no dynamical phase since the energy of dark state $\ket{D(t)}_{AB}$ is 0. Consequently, $\ket{g_{1}}_{A}\ket{g_{0}}_{B}\rightarrow\ket{g_{1}}_{A}\ket{g_{0}}_{B}$ in the second step of operation~(\ref{eq:step}) is justified.

If the system is initially in state $\ket{g_{1}}_{A}\ket{g_{2}}_{B}$, the Hamiltonian~(\ref{eq:hamiltonian1}) is modified so that the atom-cavity interaction in cavity B is involved. We assume a large detuning ($\Delta\gg g_{B},\gamma$) and take advantage of adiabatic elimination \cite{g91}. The effective Hamiltonian for atom B now reads
\begin{equation}
H_{B}=-\frac{g_{B}^{2}}{\Delta}a_{B}^{\dagger}a_{B}\ket{g_{2}}_{B}\bra{g_{2}}.
\end{equation}
This Hamiltonian can be regarded as a perturbation to Hamiltonian~(\ref{eq:hamiltonian1}). Let $\ket{D'(t)}_{AB}$ be the perturbed eigenstate of dark state~(\ref{eq:dark}) (with $g_{0}\rightarrow g_{2}$) and $E_{0}'(t)$ be its eigenenergy. The value of $E_{0}'(t)$ is no longer zero for nonzero $\Omega_{A}(t)$. After the classical pulse of Eq.~(\ref{eq:rabi}), the system also returns to its initial state $\ket{g_{1}}_{A}\ket{g_{2}}_{B}$ as in the previous case. During this operation, however, the state $\ket{D'(t)}_{AB}$ acquires a dynamical phase $\phi$ given by
\begin{equation}
\phi=\int_{t_{0}}^{t_{1}}E_{0}'(t)dt,
\end{equation}
where $(t_{1}-t_{0})$ is the operation time. The value of $\phi$ depends linearly on the width $\Delta_{t}$ of the classical pulse~(\ref{eq:rabi}) and increases as the height $\Omega_{0}$ increases. The dependency of $\phi$ on $\Omega_{0}$ can be obtained by numerical simulation.


\begin{figure}[b]
\includegraphics[width=0.45\textwidth]{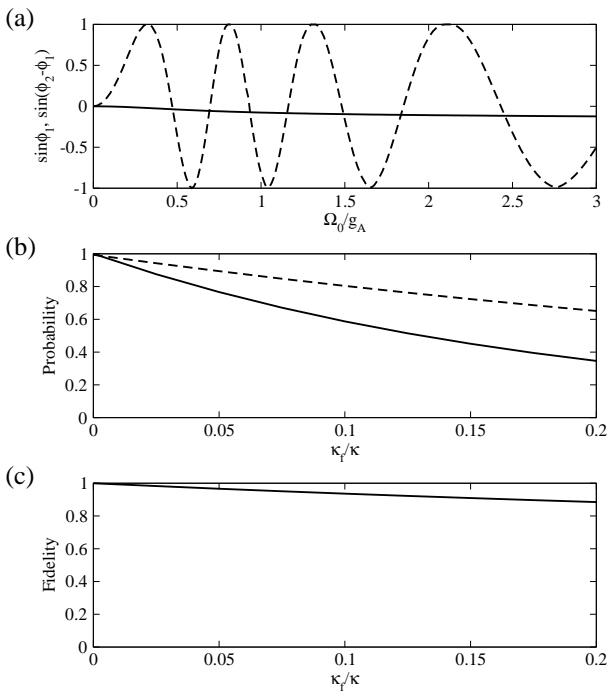}
\caption{In the short-distance regime, (a) $\sin\phi_1$ (solid curve) and $\sin(\phi_2-\phi_1)$ (dotted curve). $\phi_1$ and $\phi_2$ denote the acquired phases during the operation for the initial states $\ket{g_1}_A\ket{g_0}_B$ and $\ket{g_1}_A\ket{g_2}_B$, respectively. In case of $\Omega_0=0.465g_A$, (b) probabilities $P_1$ (solid curve) and $P_2$ (dotted curve), and (c) fidelity $F$ for an initial state $\frac12\ket{g_0}_A\ket{g_0}_B+\frac12\ket{g_0}_A\ket{g_2}_B+\frac12\ket{g_1}_A\ket{g_0}_B+\frac12\ket{g_1}_A\ket{g_2}_B$. The selected numerical parameters are $\gamma=g_{A}/10=g_{B}/10=\Delta/500=\delta\omega/7.5=\kappa$ and $\Delta_{t}=125/\kappa$.}
\label{fig:phi}
\end{figure}

We carry out numerical simulations, by directly solving the Schr{\"o}dinger equation without adiabatic approximations, for the second step of operation~(\ref{eq:step}) with a set of selected parameters: $\gamma=g_{A}/10=g_{B}/10=\Delta/500=\delta\omega/7.5=\kappa$ and $\Delta_{t}=125/\kappa$.
In the numerical simulation, we also take into account the photon loss in the fiber by introducing photon loss rate $\kappa_f$ of the fiber and adding terms $-i\sum_{n=-\infty}^{\infty}\frac{\kappa_f}{2}c_n^\dagger c_n$ in Hamiltonian~(\ref{eq:hamiltonian1}).
If a photon is lost due to the lossy fiber or the spontaneous decay of the atom, the system collapses into one of the ground states losing its phase information.
In the case of the system being collapsed into $\ket{g_1}_A$ or $\ket{g_2}_B$, it only takes the effect of lowering the fidelity of the whole operation.
If the system is collapsed into $\ket{g_2}_A$ or $\ket{g_1}_B$, however, we are faced with another problem that the third step of opertation~(\ref{eq:step}) does not transform the state into one in the qubit subspace.
In order that the resulting state be confined in the qubit subspace, we thus perform optical pumping, after the second step, by applying two classical fields corresponding to transitions $\ket{g_2}_A\leftrightarrow\ket{e}_A$ and  $\ket{g_1}_B\leftrightarrow\ket{e}_B$, which induce population transfers from $\ket{g_2}_A$ to $\ket{g_1}_A$ and from  $\ket{g_1}_B$ to $\ket{g_2}_B$, respectively.
Such an optical pumping process takes no effect when the state already exists in the desired subspace.
For the initial state $\ket{g_{1}}_{A}\ket{g_{0}}_{B}$, let $P_{1}$ be the probability that no photon is lost during the operation, and let $\phi_{1}$ be the phase that is acquired when no photon is lost. 
Let us also assign a probability $P_{2}$ and a phase $\phi_{2}$ in the same manner for the initial state $\ket{g_{1}}_{A}\ket{g_{2}}_{B}$.
The whole process is then summarized as the following operator-sum representation \cite{nc00} (omitting indices A and B for brevity):
\begin{equation}
\rho\rightarrow\sum_{i=1}^{3}M_i\rho M_i^\dagger,
\end{equation}
where
\begin{equation}
\begin{split}
M_1=&\ketbra{g_0g_0}{g_0g_0}+\ketbra{g_0g_2}{g_0g_2}\\
 &+\sqrt{P_1}e^{i\phi_1}\ketbra{g_1g_0}{g_1g_0}+\sqrt{P_2}e^{i\phi_2}\ketbra{g_1g_2}{g_1g_2},\\
M_2=&\sqrt{1-P_1}\ketbra{g_1g_0}{g_1g_0},\\
M_3=&\sqrt{1-P_2}\ketbra{g_1g_2}{g_1g_2}.\\
\end{split}
\nonumber
\end{equation}
In Fig.~\ref{fig:phi}(a), we plot $\sin\phi_1$ (solid curve) and $\sin(\phi_2-\phi_1)$ (dotted curve) with respect to $\Omega_{0}$. 
Our numerical results indicate that the acquired phases are nearly independent of the fiber loss rate $\kappa_f$ in our parametric regime.
As predicted above, the initial state $\ket{g_1}_A\ket{g_0}_B$ acquires only a small phase $\phi_1$, whereas the initial state $\ket{g_1}_A\ket{g_2}_B$ acquires a phase $\phi_2$ which increases as $\Omega_0$ increases.
The small phase change $\phi_1$ can be compensated by means of a single-qubit phase operation so that $(\phi_2-\phi_1)$ becomes the only relevant phase.
In Fig.~\ref{fig:phi}(b), we plot $P_1$ (solid curve) and $P_2$ (dotted curve) with respect to $\kappa_f$.
Here, we have chosen $\Omega_0=0.465g_A$, in which case $(\phi_2-\phi_1)$ is found to be $\pi$.
Given an initial state  $\alpha_{00}\ket{g_0g_0}+\alpha_{02}\ket{g_0g_2}+\alpha_{10}\ket{g_1g_0}+\alpha_{12}\ket{g_1g_2}$, the fidelity of the operation is easily derived as
\begin{equation}
\begin{split}
F=&\left[\left(\abs{\alpha_{00}}^2+\abs{\alpha_{02}}^2+\abs{\alpha_{10}}^2\sqrt{P_1}+\abs{\alpha_{12}}^2\sqrt{P_2}\right)^2\right.\\
&\left.+\abs{\alpha_{10}}^4(1-P_1)+\abs{\alpha_{12}}^4(1-P_2)\right]^{1/2}.
\end{split}
\label{eq:fidelity}
\end{equation}
With the same parameters as above and an initial state $\frac12\ket{g_0g_0}+\frac12\ket{g_0g_2}+\frac12\ket{g_1g_0}+\frac12\ket{g_1g_2}$, we plot in Fig.~\ref{fig:phi}(c) the fidelity $F$ with respect to $\kappa_f$.
The fidelity decreases with the fiber loss, but it remains high as long as $\kappa_f\ll\kappa$.
This observation, along with the fact that the photon loss is dominated by the spontaneous decay of the atom when $\kappa_f\ll\kappa$, leads to a conclusion that the spontaneous decay of the atom does not have critical effect since it is suppressed due to dark state evolution and the large detuning.


\section{Long-Distance Regime}

Now, we consider another regime, namely, the long-distance regime in which the time width $\Delta_{f}$ of the single photon pulse leaking out from cavity A satisfies the following condition:
\begin{equation}
\Delta_{f}<L/c.
\end{equation}
In this case, the input/output process at each cavity can be treated separately. We set the detuning parameter as $\Delta=0$ for this regime.

First, we consider the output process at cavity A. As we have considered in the previous regime, Hamiltonian $H_{A}$ has the dark state $\ket{D(t)}_{A}=\cos\theta_{A}(t)\ket{g_{1},0}_{A}-\sin\theta_{A}\ket{g_{2},1}_{A}$. If atom A is in state $\ket{g_{1}}_{A}$ initially, the population is transferred to $\ket{g_{2}}_{A}$ as $\Omega_{A}(t)$ is gradually increased from zero. During this adiabatic passage process, a single photon leaks out from the cavity. In the adiabatic limit, the pulse shape $f(t)$ of the emitted photon can be calculated analytically as \cite{dkk03}
\begin{equation}
f(t)=\sqrt{\kappa}\sin\theta_{A}(t)\exp\left[-\frac{\kappa}{2}\int_{0}^{t}\sin^{2}\theta_{A}(\tau)d\tau\right].
\label{eq:pulse}
\end{equation}

The output photon propagates through the fiber and reflects at cavity B. The input/output process at cavity B is described by the boundary condition \cite{gc85}:
\begin{equation}
c_{out}(t)=c_{in}(t)-\sqrt{\kappa} a_{B}(t),
\label{eq:boundary}
\end{equation}
and the quantum Langevin equation:
\begin{equation}
\begin{split}
\dot{b}=-i[b,H_{B}]&-[b,a_{B}^{\dagger}]\left(\frac{\kappa}{2}a_{B}-\sqrt{\kappa}c_{in}(t)\right)\\
&+[b,a_{B}]\left(\frac{\kappa}{2}a_{B}^{\dagger}-\sqrt{\kappa}c_{in}^{\dagger}(t)\right),
\end{split}
\label{eq:langevin}
\end{equation}
where $c_{in}(t)$ and $c_{out}(t)$ are the input and output field operator, respectively, and $b$ is any operator for atom-cavity system B. Let us assume that the time derivative of any operator for system B vanishes, i.e., $\dot{b}\simeq0$. This assumption is justified since the input photon pulse is generated by an adiabatic process. We also assume the strong coupling limit $g_B\gg\gamma$.

For the initial state of $\ket{g_0}_B$, the Hamiltonian reads $H_B=0$. The time derivative of $a_B(t)$ is thus derived from Eq.~(\ref{eq:langevin}) as $\dot{a_B}(t)=-\frac{\kappa}{2}a_B(t)+\sqrt{\kappa}c_{in}(t)\simeq0$. From this equation, we get $a_B(t)\simeq\frac{2}{\sqrt\kappa}c_{in}(t)$, and by substituting it into Eq.~(\ref{eq:boundary}) the relationship between the cavity input and output is derived as $c_{out}(t)\simeq-c_{in}(t)$. On the other hand, the Hamiltonian for the initial state of $\ket{g_2}_B$ reads
\begin{equation}
H_B=g_B(a_B\ket{e}_B\bra{g_2}+a_B^\dagger\ket{g_2}_B\bra{e}).
\end{equation}
In this case, we derive the time derivative of $\ket{g_2}_B\bra{e}$ as $\frac{d}{dt}(\ket{g_2}_B\bra{e})=-ia_B(\ket{g_2}_B\bra{g_2}-\ket{e}_B\bra{e})\simeq0$. Since $(\ket{g_2}_B\bra{g_2}-\ket{e}_B\bra{e})$ has a nonzero value, we come to a conclusion that $a_B(t)\simeq0$. Thus the input/output relationship reads $c_{out}(t)\simeq c_{in}(t)$. Consequently, the single photon reflected at cavity B acquires a different phase 0 or $\pi$ according to the state of atom B \cite{dk04}.

After the reflection at cavity B, the photon finally reaches cavity A. By applying an appropriate classical field of Rabi frequency $\Omega_{A}(t)$, the photon is completely absorbed in atom A and the atomic population is transferred to $\ket{g_{1}}_{A}$ by adiabatic passage \cite{l03}. Complete absorption of the photon is guaranteed if no photon is reflected during this operation, for which $\Omega_{A}(t)$ has to be adjusted to satisfy the following condition \cite{fyl00}:
\begin{equation}
-\frac{d}{dt}\ln\sin\theta_{A}(t)+\frac{d}{dt}\ln f(t)=\frac{\kappa}{2}\sin^{2}\theta_A(t).
\label{eq:matching}
\end{equation}
The different phase of 0 or $\pi$ acquired at atom B results in the different phase of the final atomic state. The resulting atomic state thus acquires a conditional phase $\phi=\pi$ as given in the second step of operation~(\ref{eq:step}). We note that the value of $\phi$ is insensitive to the particular system parameters, which is a strong point of the scheme.


\begin{figure}[t]
\includegraphics[width=0.45\textwidth]{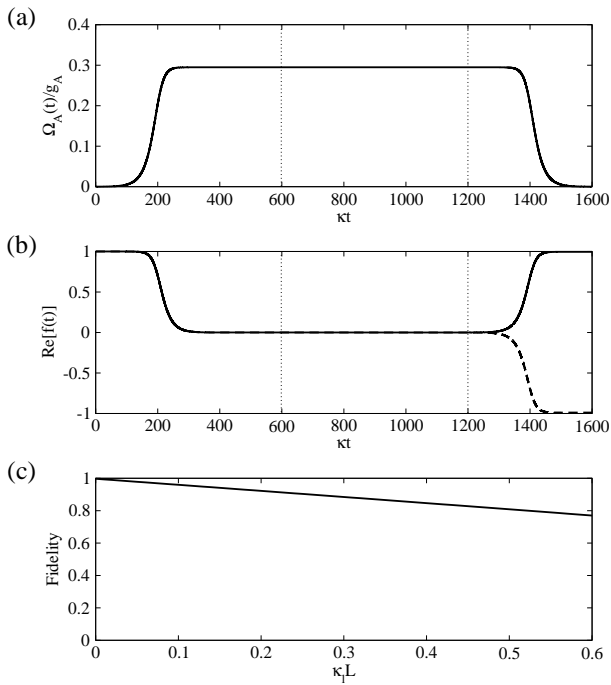}
\caption{Gate operation in the long-distance regime. (a) Rabi frequency $\Omega_{A}(t)$ of the classical pulse, (b) real parts of the amplitudes of states $\ket{g_{1}}_{A}\ket{g_{0}}_{B}$ (solid curve) and $\ket{g_{1}}_{A}\ket{g_{2}}_{B}$ (dotted curve), and (c) fidelity of the operation for an initial state $\frac12\ket{g_0}_A\ket{g_0}_B+\frac12\ket{g_0}_A\ket{g_2}_B+\frac12\ket{g_1}_A\ket{g_0}_B+\frac12\ket{g_1}_A\ket{g_2}_B$.
The selected numerical parameters are $\gamma=g_{A}/8=g_{B}/8=\kappa$, $\Delta_{t}=50/\kappa$, and $L=600\frac{c}{\kappa}$.}
\label{fig:regime2}
\end{figure}

In Fig.~\ref{fig:regime2}, we numerically demonstrate a typical gate operation process in this long-distance regime. 
The selected numerical parameters are $\gamma=g_{A}/8=g_{B}/8=\kappa$ and $\Delta_{t}=50/\kappa$. 
In Fig.~\ref{fig:regime2}(a), we plot Rabi frequency $\Omega_{A}(t)$ with respect to time. 
We set Rabi frequency $\Omega_{A}(t)$ differently from a Gaussian shape given by Eq.~(\ref{eq:rabi}) to take advantage of an analytic solution for the cavity input/output equations (\ref{eq:pulse}) and (\ref{eq:matching}) \cite{fyl00}. 
At the beginning of the gate operation, a classical field of Rabi frequency $\Omega_{A}(t)$ satisfying
$\sin\theta_{A}(t')=\sqrt{\frac{2}{\kappa\Delta_{t}}}\sqrt{\exp[2t'/\Delta_{t}]\mathrm{sech}[2t'/\Delta_{t}]}$,
where $t'=t-t_{c1}$ and $t_{c1}=200/\kappa$, is applied to atom A.
This classical field generates a cavity photon, which leaks out from the cavity with a pulse shape given by
$f(t)=\frac{1}{\sqrt{\Delta_{t}}}\mathrm{sech}[2(t-t_{c1})/\Delta_{t}]$.
The generation of the single photon pulse and the reflection of this pulse at cavity B are simulated numerically.
Here cavity B is assumed to be located at $L=600\frac{c}{\kappa}$.
In order to absorb the photon into atom-cavity A without reflection, Rabi frequency $\Omega_{A}(t)$ is finally adjusted to satisfy
$\sin\theta_{A}(t'')=\sqrt{\frac{2}{\kappa\Delta_{t}}}\sqrt{\exp[-2t''/\Delta_{t}]\mathrm{sech}[2t''/\Delta_{t}]}$,
where $t''=t-t_{c2}$ and $t_{c2}-t_{c1}=2L/c$, as shown in Fig.~\ref{fig:regime2}(a). 
We have confirmed from the numerical results that nearly no photon is emitted from cavity A during this final process.
The real parts of the amplitudes of states $\ket{g_{1}}_{A}\ket{g_{0}}_{B}$ and $\ket{g_{1}}_{A}\ket{g_{2}}_{B}$ are plotted as a solid curve and a dotted curve, respectively, in Fig.~\ref{fig:regime2}(b). 
It is clearly shown that the acquired phase corresponds to the second step of operation~(\ref{eq:step}).

The above rather idealized analysis gives the basis for the following generalized one in which the photon loss of the fiber is taken into account.
We again introduce two probabilities $P_1$ and $P_2$ as in the previous analysis for the short-distance regime.
$P_1$ and $P_2$ denote the probabilities that no photon is lost during the operation for the initial states $\ket{g_1}_A\ket{g_0}_B$ and $\ket{g_1}_A\ket{g_2}_B$, respectively.
The analysis laid out in case of the short-distance regime is applied for the current case in the same manner.
By means of the same sort of optical pumping, the state can be confined in the qubit subspace, and the fidelity is also given by Eq.~(\ref{eq:fidelity}).
The above numerical simulation gives $P_1$ and $P_2$ in case of no photon loss in the fiber.
The values are found to be $P_1=0.992$ and $P_2=0.977$, which gives the fidelity $F=0.997$ when the initial state is chosen as $\frac12\ket{g_0}_A\ket{g_0}_B+\frac12\ket{g_0}_A\ket{g_2}_B+\frac12\ket{g_1}_A\ket{g_0}_B+\frac12\ket{g_1}_A\ket{g_2}_B$.
If the fiber is not lossless and the photon loss rate per unit length of the fiber is denoted as $\kappa_l$, we get modified probabilities $P_1\rightarrow (1-\kappa_l L)^2P_1$ and $P_2\rightarrow (1-\kappa_l L)^2P_2$ for the two initial states, respectively.
By substituting these probabilities into Eq.~(\ref{eq:fidelity}) and choosing the initial state as $\frac12\ket{g_0}_A\ket{g_0}_B+\frac12\ket{g_0}_A\ket{g_2}_B+\frac12\ket{g_1}_A\ket{g_0}_B+\frac12\ket{g_1}_A\ket{g_2}_B$, we plot in Fig.~\ref{fig:regime2}(c) the fidelity $F$ of the operation with respect to $\kappa_l$.
As in the short-distance regime, the numerical results show that the gate works faithfully when the photon loss in the fiber is small, namely $\kappa_l L\ll 1$, and the spontaneous decay of the atom does not have critical effect.


\section{Summary}

In summary, we have shown that a two-qubit controlled unitary operation between two distant atoms is allowed by simply connecting them through a quantum transmission line. 
We have analyzed it in two regimes, namely, the short-distance regime and the long-distance regime.
The scheme is based on the adiabatic passage and the cavity QED interaction. 
Provided a single photon is reliably transmitted between two cavities, the gate works with high fidelity due to the inherent resistance of the adiabatic evolution against spontaneous decay.
Since our scheme is much simpler than other indirect methods, it is expected to improve the scalability of quantum computers based on trapped neutral atoms or ions.

\begin{acknowledgments}
This research was supported by the "Single Quantum-Based Metrology in Nanoscale" project of the Korea Research Institute of Standards and Science.
\end{acknowledgments}


\end{document}